\documentclass[sigconf]{acmart}

\usepackage{booktabs}
\usepackage{graphicx}
\usepackage{balance}
\usepackage{caption}
\usepackage{threeparttable}
\usepackage{tabularx}
\usepackage{colortbl}
\usepackage[font=small,skip=4pt]{caption}

\usepackage{subcaption}
\usepackage{amsmath}
\usepackage{stmaryrd}
\usepackage{algorithm}
\usepackage{alltt}
\usepackage{amsfonts}
\usepackage{amssymb}
\usepackage{balance}
\usepackage{color}

\begin{document}

\title{Addressing the Free-Rider Problem in Public Transport Systems}

\author{Vaibhav Kulkarni, Bertil Chapuis,\\ Beno\^{i}t Garbinato}
\affiliation{%
  \institution{UNIL-HEC Lausanne}
}
\email{firstname.lastname@unil.ch}

\author{Abhijit Mahalunkar}
\affiliation{%
  \institution{School of Computing\\ DIT-Dublin}
}
\email{abhijit.mahalunkar@mydit.ie}

\renewcommand{\shortauthors}{V. Kulkarni et al.}

\begin{abstract}

Public transport network constitutes for an indispensable part of a city by providing mobility services to the general masses. 
To improve ease of access and reduce infrastructural investments, public transport authorities often adopt {\bf{proof of payment}} system. 
Such a system operates by eliminating ticket controls when boarding the vehicle and subjecting the travelers to random ticket checks by affiliated personnel (controllers). 
Although cost efficient, such a system promotes {\bf{free-riders}}, who deliberately decide to evade fares for the transport service.
A recent survey by the association of European transport, estimates hefty income losses due to fare evasion, highlighting that free-riding is a serious problem that needs immediate attention.
To this end, we highlight the attack vectors which can be exploited by free-riders by analyzing the crowdsourced data about the control-locations. 
Next, we propose a framework to generate {\bf{randomized control-location traces}} by using generative adversarial networks (GANs) in order to minimize the attack vectors.
Finally, we propose metrics to evaluate such a system, quantified in terms of increased risk and higher probability of being subjected to control checks across the city. 

\end{abstract}



\maketitle

\section{Introduction}\label{sec:introduction}

Public transport free-riding is raising alarming concerns due to the serious income losses (EUR 250 million in Switzerland, EUR 80 million in Paris and EUR 120 million in Berlin) to the public transport service authorities~\cite{swiss}. 
Switzerland incurs a financial damage in terms of a revenue loss between 1 to 15\% (mean of 5\%), estimated to be one of the highest in the world~\cite{furst2012free}.
The increasing fines have given rise to insurance organizations, which promote fare-dodging by covering the fines for free riding~\cite{planka}.
The cost-benefit analysis conducted by several public transport authorities show that the installation of a barrier separated systems or turnstiles to be unreasonable~\cite{furst2012free}. 
In addition to causing financial losses, free-riding undermines the integrity of honest passengers who need to pay more for the service in the subsequent years due to fare increments.  
Therefore free evasion needs immediate attention by devising efficient combat strategies by anticipating the attack vectors available to the free-riders.

To this end, we propose a framework for public transport control, with the objective of increasing the risk involved in traveling without a valid ticket.  
A project with a similar objective was undertaken in New York City called project Eagle~\cite{eagle}. 
The eagle team located {\bf{hotspots}}, where increased fare evasion activity was taking place and heightened the frequency of controls at these places by increasing the number of team members.
However, this led to increased expenses due to the appointment of additional personnel and elevated free-riding activity in the other areas of the city.  
Therefore, a satisfactory tradeoff needs to be maintained between the expenses incurred and the control efficacy.

As a first step towards addressing this problem, we study the current state of public transport controlling by analyzing the crowdsourced data about the locations of ticket controllers collected in Lausanne, Switzerland.
We highlight several attack vectors, resulting primarily due to the distinct predictability and lack of randomness in the movement patterns of the controllers. 
To this end, we propose a framework to generate control locations and the routes connecting them to maximize the informativeness about free-riding activity. 
The location set and the routes can be determined by analyzing the population density flow in the city by gathering the ridership data collected from the public transport authorities.
This data is used to train a GAN~\cite{Goodfellow2014GenerativeAN}, which is used to formulate control-location traces satisfying certain cost and informativeness bounds. 
Finally, we suggest statistical metrics to evaluate the generated traces based on their distribution similarity and randomness. 

{\noindent{\bf{Problem Statement.}}} Given a control-location $p$, having an associated cost (transit time), $C(p)\mid p\in L$, where $L$ is the set of all the city control-locations.   
The route, $R$ between two control-locations, $p,q\in L$ has an associated cost, $C(R)$.
Let $T$ be a trace expressed as a sequence of these routes with a cost, $(C(T))$
In addition, every location and route has an associated quality, i.e. informativeness (\#travelers controlled) denoted as $Q(T)$.  
Our goal is to generate control-location traces such that $\max Q(T) \mid C(T)<B, L ~ D(\mu_{r}, \sigma)$, where the cost is bounded by $B$, $D$ is the distribution of the locations and routes based on the ridership and the randomness is quantified in terms of $\mu_{r}$.

\begin{figure*}[t!]
    \centering
    \begin{subfigure}[b]{0.337\textwidth}
        \includegraphics[width=\textwidth]{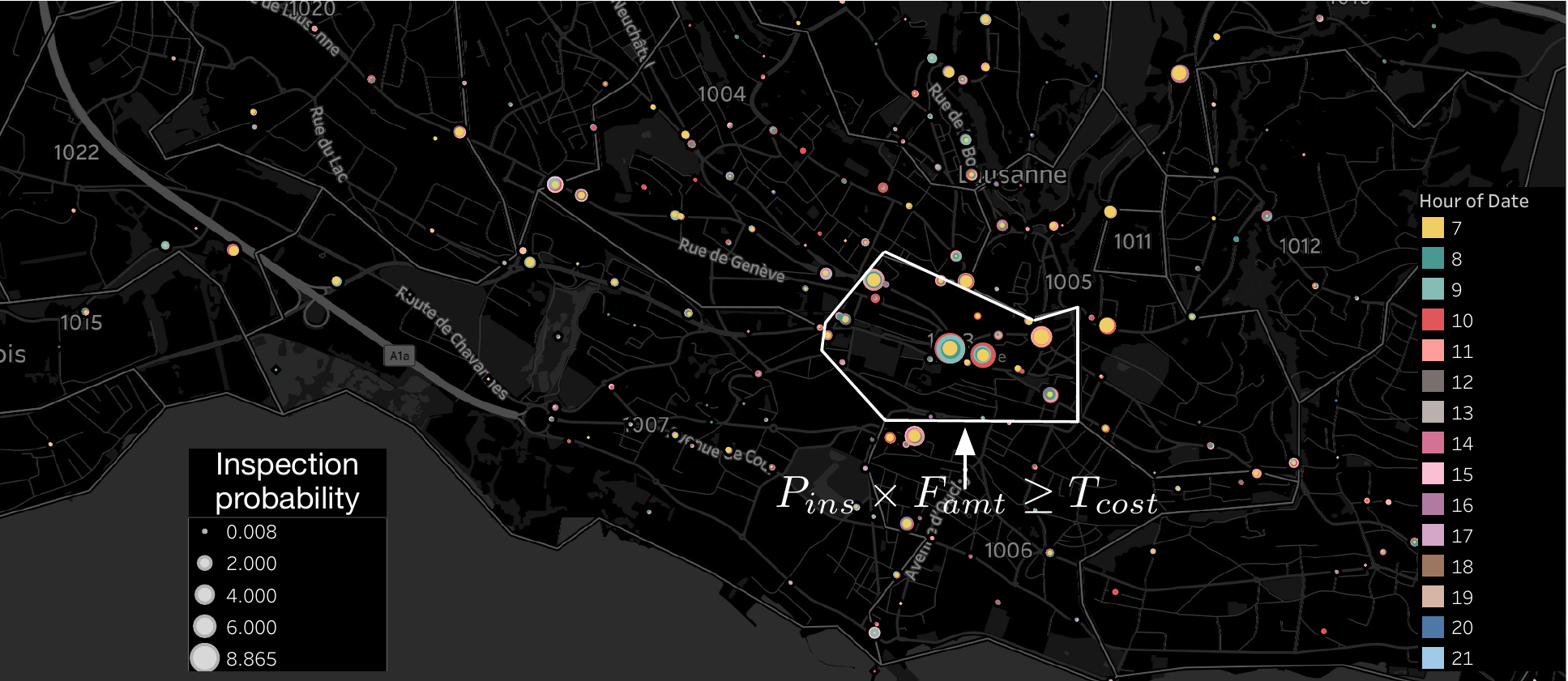}
        \caption{Inspection probability according to zip-codes}
        \label{fig:p1}
    \end{subfigure}
    \begin{subfigure}[b]{0.318\textwidth}
        \includegraphics[width=\textwidth]{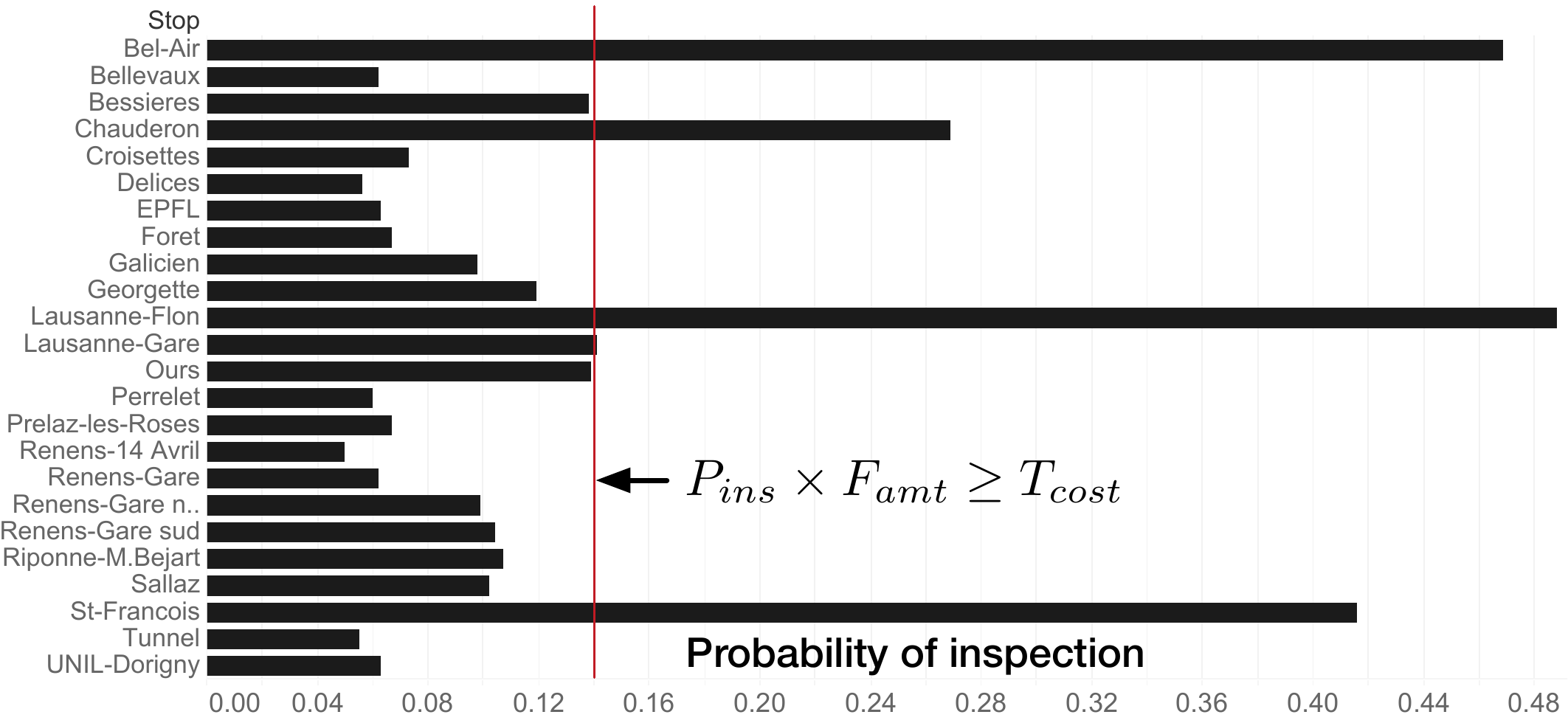}
        \caption{stops v/s. inspection probability}
        \label{fig:p2}
    \end{subfigure}
    \begin{subfigure}[b]{0.337\textwidth}
        \includegraphics[width=\textwidth]{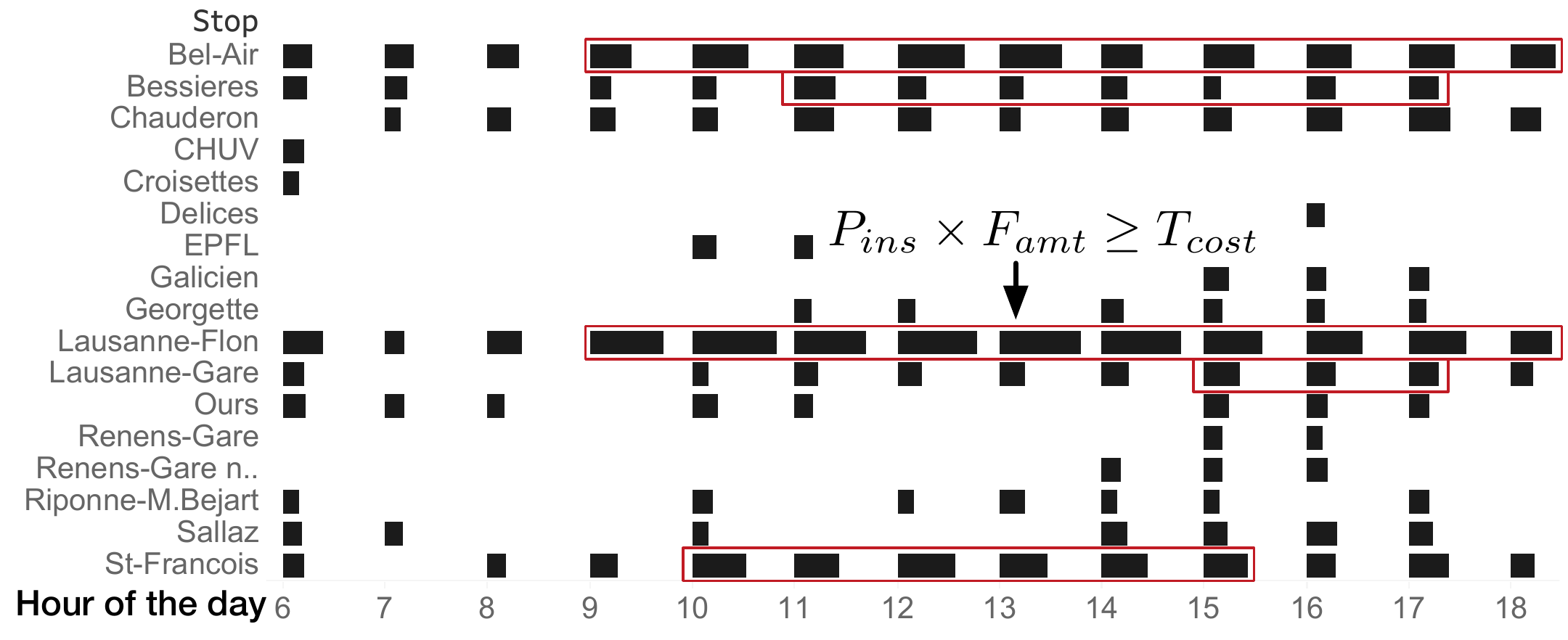}
        \caption{Stops v/s. hour v/s. inspection occurrences}
        \label{fig:p3}
    \end{subfigure}   
    	\caption{Analysis of the crowdsourced data regarding locations of ticket controllers for over four years.}
    \label{fig:p123}    
\end{figure*}

\section{Attack Vectors}
\label{sec:atk_surfaces}

We determine the attack vectors by analyzing the crowdsourced data gathered via an application{\footnote{Busted App: busted-app.com}} which allows the users to mark the current controller's location.    
We collect a total of 14,500 location coordinates spanning over four years, regarding the observations of controller movements in the city of Lausanne, Switzerland contributed by more than 20,000 users. 
We compute: (1) the probability of getting inspected at each station/on each line in the city, (2) correlations between location and time and (3) controller movement predictability.

As shown in Figure~\ref{fig:p1} and~\ref{fig:p2}, the control frequency is noticeably higher at a select few locations as compared to the rest.      
Furthermore, we also observe strong correlations between certain locations and their inspection times~\ref{fig:p3}, locations and the control position (i.e. inside or outside the vehicle).
In addition, we detect recurring patterns in the controlling schedules across weekday/weekend/seasonal trends and high movement prediction confidence.

This information can be be used to devise attacks such as selective ticket purchasing, i.e. purchase a ticket only if the inspection probability of the ride ($P_{ins}$) $\times$ the fine amount ($F_{amt}$) is greater than the price of the ticket ($T_{cost}$).  
It can also facilitate other attacks such as avoiding stops where controlling is performed outside the vehicle, path re-routing to minimize the inspection probability and exploiting the trends to predict and avoid the future control locations.

\section{Control-Trace Generation}
\label{sec:cluster_extract}

In this work, we focus on minimizing the attack vectors by eliminating the non-uniform control probability distribution and the spatiotemporal correlations. 
However, unconditional random sampling from the available location set may weaken/increase control at key/minimal-activity locations.  
Our solution to generate the randomized control traces, therefore lies in sampling from the true city-wide population density flows.

As the controllers also need to traverse between the locations, we first model the city's public transport network as a bipartite graph.
Here, the stations and the routes are represented by nodes~\cite{Ferber2008PublicTN} and the edges denote the station nodes serviced by a route nodes. 
Such a graph, facilitates quantifying the path cost $C$ and the quality $Q$ by utilizing the nodes connection degree (neighborhood) and the population density flow to result in a time varying network~\cite{Galati2013AnalyzingTM}.

\begin{figure}[!h]
\centering
\includegraphics[scale=0.36]{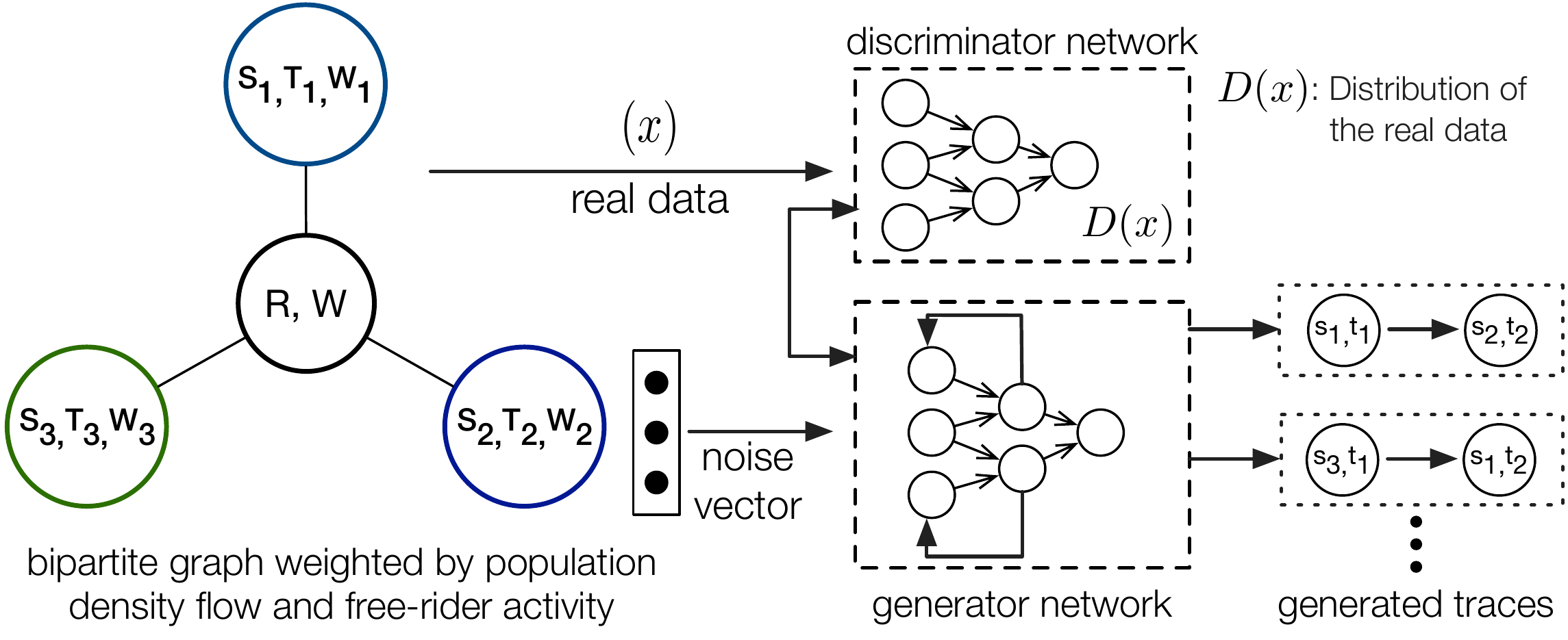}
\caption{Model overview to generate randomized control spots}
\label{fig:p5}
\vspace{-15px}
\end{figure}

Next, we train a Sequence GAN~\cite{Yu2017SeqGANSG} by using the sequences derived from the time varying network, ordered with respect to the quality and cost criteria as depicted in Figure~\ref{fig:p5}.  
We rely on GANs in this context due to their ability to accurately capture a parent data distribution which can be used to generate synthetic data. 
Thus the generated control-location traces satisfy the distribution of the population density flow and are significantly different to each other.   
This accounts for the required randomness as well as adherence to the essential criteria's of controlling.  
\vspace{-4px}   

\section{Conclusion \& Future Work}
\label{sec:conclusion}

In this work, we have highlighted some of the factors resulting in the failure of current public transport controlling strategies, which facilitates free-rider activity.     
We have proposed a framework to address this issue by generating control-location traces that are spatiotemporally randomized but adhere to some of the key requirements of controlling.   
Our future work aims at training the model with larger data volumes to incorporate the seasonal trends at a higher granularity.
We will conduct a detailed evaluation of the generated traces with respect to the distribution similarity in terms of the Kolmogorov-Smirnov test and randomness in terms of the root mean square error.
We also plan to practically validate our approach by comparing the free-rider activity before and after adopting our approach. 

\footnotesize{
\bibliographystyle{ACM-Reference-Format}
\bibliography{sample-sigconf.bbl}
}

\end{document}